\documentclass[epjST]{svjour}
\usepackage{graphics}
\begin{document}
\title{Quantum Annealing - Foundations and Frontiers}

\author{Eliahu Cohen\inst{1}\fnmsep\thanks{\email{eliahuco@post.tau.ac.il}} \and Boaz Tamir\inst{2}}
\institute{School of Physics and Astronomy, Tel Aviv University, Tel Aviv, Israel \and Faculty of Interdisciplinary Studies, Bar-Ilan University, Ramat-Gan, Israel}
\abstract{
We briefly review various computational methods for the solution of optimization problems. First, several classical methods such as Metropolis algorithm and simulated annealing are discussed. We continue with a description of quantum methods, namely adiabatic quantum computation and quantum annealing. Next, the new D-Wave computer and the recent progress in the field claimed by the D-Wave group are discussed. We present a set of criteria which could help in testing the quantum features of these computers. We conclude with a list of considerations with regard to future research.
} 

\maketitle

\section{Introduction}
\label{ECBTintro}
Finding the best solution out of all feasible solutions as fast as possible. This is not only a major goal for engineers, computer scientists and other researchers, but rather our usual aim in everyday life.
For more than 70 years, computers have been used to solve optimization problems. While most problems were essentially unchanged, computers and optimization techniques have rapidly developed. Gigantic computers composed of numerous
vacuum tubes and diodes were soon replaced by tiny silicon chips, and simple Monte Carlo (MC) methods gave place to adaptive simulated annealing algorithms. During the last two decades, a great deal of attention has focused on quantum computation following a sequence of results \cite{Shor,Grover} suggesting that quantum computers are more powerful than
classical probabilistic computers. Following Shor's result \cite{Shor} that factoring and extraction
of discrete logarithms are both solvable on quantum computers in polynomial
time, it is natural to ask whether other non-deterministic polynomial (NP) problems
can be efficiently solved on quantum computers in polynomial time.
It was Feynman's idea that quantum phenomena could not always be simulated
by classical computers, and whenever there are such simulations there is an exponential
growth in the resources needed.
On 1982 Feynman conjectured that quantum computers can be programmed to
simulate any local quantum system \cite{Feynman}. Since then, a vast literature has been written,
addressing the theoretical and practical advantages and difficulties. On 1996 Lloyd
supported Feynman's claim and concluded \cite{Lloyd}: ``The wide variety of atomic, molecular
and semiconductor quantum devices available suggests that quantum simulation
may soon be reality''. Just 3 years later, D-Wave systems were founded with the goal
of making practical quantum computers \cite{Dwave}.
Indeed, quantum technology is maturing to the point where quantum devices,
such as quantum communication systems, quantum random number generators and
quantum simulators are built with capabilities exceeding classical computers.
Quantum annealers \cite{Das}, in particular, solve hard optimization problems by evolving a known initial configuration at non-zero temperature towards the ground state of a
Hamiltonian encoding a given problem. Quantum annealing is a generalization of classical simulated annealing \cite{Laarhoven}, an approach to solve optimization problems based on the observation that the problem's cost function can be viewed as the energy of a physical system, and that energy barriers can be crossed by thermal hopping. However, to escape local minima it can be advantageous to explore low energy configurations quantum mechanically
by exploiting superpositions and tunneling. Quantum annealing and adiabatic
quantum computation are algorithms based on this idea, and programmable
quantum annealers, such as the D-Wave computers, are its physical realization.
Quantum information processing offers dramatic speed-ups, yet is famously susceptible
to decoherence, the process whereby quantum superposition decays into
mutually exclusive classical alternatives, a mixed state, thus robbing quantum
computers of their power. For this reason, many researchers put in question the quantum features of the D-Wave computers \cite{Boixo1,Boixo2,Aaronson}. This work discusses the controversy concerning the properties of the D-Wave computers.

The work's outline is as follows: we discuss in Sec. \ref{ECBTsec:1} classical optimization methods such as Monte-Carlo, Metropolis and simulated annealing. In Sec. \ref{ECBTsec:2} we briefly overview quantum computers and algorithms. In Secs. \ref{ECBTsec:3} and \ref{ECBTsec:4} we discuss quantum adiabatic computation and quantum annealing respectively. Sec. \ref{ECBTsec:5} reviews Josephson Junctions (JJs) and presents the physical circuits which  create the flux qubits in D-Wave's computers. Sec. \ref{ECBTsec:6} is a discussion of the main characteristics of the D-Wave computer, as well as future implementations of quantum annealers. (For a complementary discussion see \cite{Cohen}).

\section{Simulated Annealing and Other Classical Optimization Methods}
\label{ECBTsec:1}
Before we reach quantum annealing, it might be instructive to briefly review the methods and ideas which led to its development.
\subsection{Monte-Carlo}
\label{ECBTsec:1.1}
The Monte-Carlo (MC) method \cite{MC1,MC2,MC3,Tee} consists of solving various computational problems by means of constructing some random process for each such problem, where the statistical
properties of the process equal the required quantities. These quantities are then
determined approximately by means of sampling. MC methods stand for a very broad family of stochastic techniques being
used in various fields such as statistical physics, astrophysics, nuclear physics and
QCD, as well as computational biology, economics, traffic flows and VLSI design \cite{MC1,MC2,MC3}.
The reason for this broad use of MC methods is they allow examining multivariable
systems that we otherwise cannot. Solving equations which describe the
interactions between two atoms is fairly simple; solving the same equations for
hundreds or thousands of atoms is impossible. With MC methods, a large system can be sampled a multitude of times in its various configurations, and that data can be
used to describe the system as a whole.
In the general case, one uses various distributions of random numbers, each distribution reflecting a particular process in a sequence of processes such as the
diffusion of neutrons in various materials, in order to finally simulate samples that
approximate the real history of the desired process.\\

MC methods are based on Markov stochastic processes. The relaxation time of a Markov chain to its stationary distribution state is governed by the reciprocal of the difference between the two highest eigenvalues of the Markov matrix \cite{Levin}. We will see in Sec. \ref{ECBTsec:3} the analogous ``relaxation'' time in adiabatic computation.

\subsection{The Metropolis algorithm}
\label{ECBTsec:1.2}
A common application of MC method is carried out by the Metropolis
algorithm \cite{Metropolis} (also called the Metropolis-Hastings algorithm). The algorithm
belongs to an important class of MC methods known as: Markov Chain MC. These
simulate a probability distribution function by constructing a Markov (memoryless)
chain that has the desired distribution as its equilibrium distribution.

As an example we take a two dimensional lattice composed of $N$ classically
interacting molecules with a distance $d_{ij}$ between molecules $i$ and $j$. The algorithm will simulate a Boltzmann distribution and will allow the computation of expectations
of functions under this distribution. Within the canonical ensemble, the equilibrium
property of any quantity $A$ is calculated according to:
\begin{eqnarray}\label{ECBT_Abar}
\bar{A}=\frac{\int Aexp(-E/k_bT)d^{2N}pd^{2N}q}{\int exp(-E/k_bT)d^{2N}pd^{2N}q},
\end{eqnarray}
where $k_b$ is Boltzmann constant, $T$ is the temperature, $d^{2N}pd^{2N}q$ is a volume element in the $4N$ dimensional phase space and $E$ is assumed to be a potential energy of the form:
\begin{eqnarray}
E=\sum_{i=1}^{N}\sum_{j\neq i}V(d_{ij}),
\end{eqnarray}
where $V$ is some external potential depending only on the distance between the molecules.

The algorithm steps are as follows: \\
(1) $N$ particles are placed in a two-dimensional lattice. \\
(2) Each of the particles is moved in succession according to:
\begin{eqnarray}
X \rightarrow X + r\delta_x~,~Y \rightarrow Y + r\delta_y
\end{eqnarray}
where $r$ is the maximum allowed displacement, and $\delta_x$,$\delta_y$ are random numbers between $-1$ and $1$, {\it i.e.} the choice of the next position is made using a uniformly distribution on a square of side $2r$ centered about the particle's original position, this is known as the ``proposal distribution''. \\
(3) The change in energy $\Delta E$, which is caused by the movement is then calculated.\\
(4a) If $\Delta E\le 0$, which means that the new state is energetically preferable, the particle will stay in its new position with probability $1$. \\
(4b) If $\Delta E > 0$ the particle will stay in its new position with probability $exp(-\Delta E/k_bT)$, that is, another random number $0 \le\delta\le 1$ will be generated
and compared to $exp(-\Delta E/k_bT)$. If $\delta \le exp(-\Delta E/k_bT)$, the particle will stay in its new position. Else, the particle will be returned to its previous position.\\

The probability distribution used in step (4b) is known as the ``acceptance distribution''. Note the similarity of the ``acceptance distribution'' to the ``target distribution'' which is the Boltzmann distribution. Given the ``proposal distribution'' and the ``acceptance distribution'' there are simple sufficient conditions to guarantee the existence of a stationary target distribution, these are known as the ``detailed balance'' conditions. Step 4b above guarantees the fulfillment of the detailed balance condition. The above process will produce a set of points on the lattice that
are distributed according to the Boltzmann distribution.

\begin{figure}
\center
\resizebox{0.25\columnwidth}{!}{
 \includegraphics{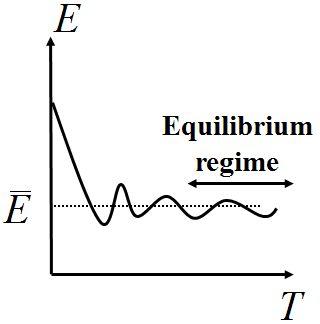} }
\caption{Typical results of transition to equilibrium}
\label{ECBTfig1}
\end{figure}

This method enables to approximately find $\bar E$, as well as any other average property in equilibrium $\bar A$ described by Eq. $\ref{ECBT_Abar}$ above (see Fig. \ref{ECBTfig1}).

Such algorithms have several disadvantages, first, neighboring points (with respect to the order of points coming out of the algorithm) are clearly correlated, so one has to sample from a sub series, second, the initial segment of points are distributed far from the target distribution.

Note that the temperature is constant throughout the algorithm. A gradual change in the temperature would imply the use of the simulated annealing algorithm.
\subsection{Simulated Annealing}
\label{ECBTsec:1.3}

Kirkpatrick {\it et al.} \cite{Kirkpatrick} and independently $\check{C}ern\acute{y}$ \cite{Cerny} realized that there exists a profound analogy
between minimizing the cost function of a combinatorial optimization problem and
the slow cooling of a solid until it reaches its low energy ground state. Since then, the research into this algorithm and its applications have
evolved into a field of study in its own.
By substituting cost for energy and by executing the Metropolis algorithm at a
sequence of slowly decreasing temperature values, Kirkpatrick and his co-workers
obtained a combinatorial optimization algorithm, which they called ``simulated
annealing''. In condensed matter physics, annealing denotes a physical process during which a solid in a heat bath is heated up by increasing the temperature of the bath to a maximum value at which all particles of the solid randomly arrange themselves in a liquid phase, followed by cooling through slowly lowering the temperature of the bath. In this way, all particles arrange themselves in the low energy ground state of a
corresponding lattice, provided the maximum temperature is sufficiently high and the
cooling is carried out sufficiently slowly.
The Metropolis algorithm can be used to generate such a sequence of configurations. Let the acceptance distribution be:
\begin{eqnarray}
\frac{1}{Z(c)} e^{-\Delta(C)/c}
\end{eqnarray}
\noindent where the cost function $C$ and the control parameter $c$ take the roles of energy and temperature, respectively and $Z(c)$ is some normalizing factor. The simulated annealing algorithm can now be viewed as sequential applications of the Metropolis algorithm with decreasing values of the control parameter $c$ (this is the analog of cooling a metal during the annealing process). Initially, the control parameter is given a high value, then it is gradually lowered according to some annealing schedule determined by the user. The specific way of lowering the control parameter is case dependent, but it usually starts with some high value and decreases to $0$ towards the end of time allocated for running the algorithm. By the well known theorem of Geman \cite{Geman} if we lower the control (temperature) parameter logarithmically with time: $c(t)=\alpha N /log(t)$, (where $\alpha$ is some proportionality factor), we guarantee the convergence to the solution (see also Hajek theorem \cite{Bertsimas}). This could take an infinite time, hence the algorithm is terminated at some final time $t_f$  for which virtually no deteriorations are accepted anymore. If we stop the algorithm when the temperature is relaxed to $c(t_f)=\delta$ then the required time is $ t_f=e^{\frac{\alpha N }{\delta}}$ (which will later be compared with the time required in a quantum annealing scheme). The final configuration is then taken as the solution of the problem at hand.
A recent application of simulated annealing for restoration of images and video signals was presented in \cite{Cohen2,Cohen3}.
Simulated annealing has a major drawback which is its sensitivity to clustering. If
the ``wiring'' of the computer is such that at some point in the evolution a cluster is
formed, then the evolution is slowed down \cite{Laarhoven}. In addition, it can not simplify known hard optimization problems; Barahona proved that finding the global minimum of both the $2D$ and $3D$ Ising systems is NP-hard \cite{Barahona}. For this reason it is suggested to improve this model by using the computational power of a quantum annealer.

Several variants of the simulated annealing algorithm, which adjust automatically the model's parameters were long ago developed. Two of these are Adaptive Simulated Annealing \cite{Ingber1,Ingber2}, and Simulated Quenching \cite{Vasana}.

To conclude this section we note that there are other metaheuristic methods that use different approaches of solving optimization problems, such as Genetic Algorithm \cite{Holland}, Hill Climbing \cite{Lin}, Simulated Evolution, \cite{Fogel},  Stochastic Approximations \cite{Robbins} and many other models. Here we focus on simulated annealing which directly lead us to quantum annealing to be thoroughly discussed in the rest of the work.  As we shall see, quantum fluctuations, as well as tunneling, will take the role of thermal fluctuations.

\section{Quantum Computers and Quantum Algorithms}
\label{ECBTsec:2}
Quantum computation was first suggested by Feynman as a way to overcome the
problem of simulating quantum phenomena on a classical computer \cite{Feynman}. Feynman
pointed out that a set of measurements on EPR entangled quantum particles could
not be simulated in principle by classical means. Moreover, even when one can use classical
computers to simulate quantum phenomena the growth in resources is exponential.
Therefore the natural way is to think of quantum computers. Soon Benioff \cite{Benioff} and Deutsch \cite{Deutsch1} presented a quantum version of a Turing machine (see also \cite{Albert}).
However, the quantum Turing machine model was not practical. In 1989 Deutsch
came with the idea of a quantum gate network computer \cite{Deutsch2}. He also provided a strong
argument showing that any finite dimensional unitary operator on a quantum state
could be simulated by a simple universal gate. This universal gate approximates any
other quantum gate by using the well-known Kronecker \cite{Kronecker} approximation. Deutsch
also presented the first known ``quantum algorithm'' later extended to the
Deutsch-Josza algorithm \cite{Jozsa}. These algorithms can distinguish between a balanced
function and a constant one by using a small number of measurements. They showed
an exponential benefit over classical deterministic algorithms.
Quantum computers have no architecture and in that sense they resemble old,
one purpose, analogue computers. Since the work of Deutsch, two main
families of algorithms were presented: in 1996 Grover \cite{Grover}  presented a quantum search
algorithm for an element in an unsorted array. The Grover algorithm has a speed up
of a square root over the classical search algorithm. Although such a speed up does
not cross a complexity class line, it shows a clear (and proven) gap between the
quantum and classical computational complexity. The Grover algorithm can be
generalized and used to speed-up many classical algorithms \cite{Boyer}. Later, Grover also used
the above algorithm to present a scheme for the construction of any superposition in
an $N$ dimensional vector space using at maximum $\sqrt{N}$ steps \cite{Grover1}.
In 1994, Shor \cite{Shor1} presented a polynomial algorithm for prime factorization and
discrete logarithms. So far all known classical algorithm for the factorization or
discrete logarithm have exponential complexity. Therefore the reduction in complexity
looks very good. However we have no proof for the claim that the complexity
of such classical algorithms should be bounded from below by exponential function.
Shor's algorithm used a Fourier transform module that can identify a hidden symmetry.
This was the extension of a previous quantum algorithm by Simon \cite{Simon}. The
Fourier quantum module could serve also for phase estimation \cite{Kitaev}, for order finding \cite{Shor1,Shor97},
and in general for the identification of other ``Hidden Subgroup'' of symmetries \cite{Jozsa1}.
A severe drawback in quantum computation was and still is the problem of
decoherence \cite{Namiki}. It is hard to construct a stable superposition of even a small number of
qubits. It is even harder to apply unitary gates between the qubits. So far there are
several suggestions as to the way to construct a quantum computer. Clearly, it is
enough to construct the set of universal quantum gates (for the existence of such a set see \cite{DiVincenzo1}): a $XOR$ or what is known in quantum
computation as a $CNOT$ gate, and a one qubit rotation.
A major breakthrough came with the presentation of fault tolerant quantum
gates \cite{Shor2}. These and error correction quantum codes sustain the hope that one day a quantum computer could indeed come true.

In 2000, Farhi \cite{Farhi} described a new model of quantum computation based on the
quantum adiabatic theorem. It turned out that the quantum adiabatic model is
equivalent to the quantum gate network model of Deutsch \cite{Aharonovd}. The adiabatic model is discussed in the next section.

Several criteria were suggested by DiVincenzo \cite{DiVincenzo2} for the physical possibility of the realization of quantum computers: \hfill\break
(1) One clearly needs a physical representation of qubits. \hfill\break
(2) The coherence time of the qubits should be large enough to allow the computation.\hfill\break
(3) There should be a physical mechanism realizing the unitary evolution of the
qubits. This mechanism must be controlled. \hfill\break
(4) Initial qubit states should be conveniently prepared. \hfill\break
(5) There should be a way performing a projective measurement of the final qubit
states. \hfill\break
We shall go back to the above requirements when discussing the qubit circuits of the D-Wave computer.

Small scale quantum computers based on different kinds of physical qubits have
been implemented so far. To name just a few: single photon quantum computers \cite{Chuang,Knill},
nuclear spins \cite{Nuclear,Vandersypen}, trapped ions \cite{Cirac}, neutral atoms in optical lattices \cite{Brennen}, states of superconducting circuits \cite{Mooij}, quantum dots \cite{Imamog} and electrons' spin on Helium \cite{Platzman}. We will focus on quantum adiabatic computation as the ``software'' and on superconducting flux qubits as the ``hardware''.

\section{Quantum Adiabatic Computation}
\label{ECBTsec:3}
Adiabatic quantum computation (AQC) is a scheme of quantum computation that is theoretically predicted to be more robust against noise than most of the above methods \cite{Childs,Paz,SLloyd}. In this scheme a physical system, is initially prepared in its known lowest energy configuration, or ground state. The computation involves gradually deforming the system's Hamiltonian, very slowly, to assure that the system remains in its ground state throughout the evolution process. One designs the evolution of the Hamiltonian such that the ground state of the final Hamiltonian is the solution to the optimization problem. AQC is based on the adiabatic theorem stated by Born and Fock \cite{Fock}:

``A physical system remains in its instantaneous eigenstate if a given perturbation is acting on it slowly enough and if there is a gap between the eigenvalues and the rest of the Hamiltonian spectrum.''

The Hamiltonian is therefore time dependent $H=H(t)$. The initial Hamiltonian $H(0)=H_0$ and its lowest energy eigenvector should be easy to construct. We assume the final Hamiltonian $H_T$ is also easy to construct, however its $0$ energy eigenvector is the solution to our optimization problem and could be hard (NP) to find with a classical algorithm. The Hamiltonian of the AQC is therefore:
\begin{eqnarray}
H(t) = \frac{t}{\tau} H_0 + (1- \frac{t}{\tau}) H_T = sH_0 +(1-s) H_T
\end{eqnarray}
The complexity of the adiabatic algorithm is manifested in the time it takes to evolve the computer from its initial values to its final state. It can be shown that the adiabatic approximation is valid when the ``annealing'' time $\tau$ satisfies:
\begin{eqnarray} \label{ECBTadiabaticT}
\tau>> \frac {\max_{0\le s \le 1}[ \langle 1(s)| \frac{dH(s)}{ds} |0(s)\rangle]}{\min_{0\le s \le 1}[ \Delta_{10}(s)]^2}
\end{eqnarray}
where $|{i(s)}\rangle$ for $i=0,1$ are the ground and first excited states of $H(s)$, and $\Delta_{10}(s)$ is their energy difference. The logic behind this formula is as follows: we want to bound from below the difference between the two lowest eigenvalues. The minimal time of evolution should be proportional to that bound. However, we cannot increase this bound by artificially blowing up the Hamiltonian itself, therefore we need the above quotient. If this first gap decreases very rapidly (exponentially) as a function of the number of variables in our problem then we are in trouble, we should slow down the evolution (exponentially).

AQC was first suggested by Farhi {\it et al.} \cite{Farhi}, where the 3-SAT problem was discussed. Farhi's computer does not solve NP problems in polynomial time. It turns out that the time complexity of such algorithms is hard to compute and so far known to be exponential. He also suggested, that for special cases, one can reduce the time complexity by finding a tensor decomposition for the whole Hamiltonian and using this to show that the time complexity is low.

Later on, Aharonov {\it et al.} showed \cite{Aharonovd} that adiabatic quantum computation is equivalent to the Deutsch circuit model of quantum computation. The equivalence between the models provides a new vantage point from which to tackle the central issues in quantum computation, namely designing new quantum algorithms and constructing fault tolerant quantum computers. Unfortunately, the proof in \cite{Aharonovd} does not provide a simple way to go from one model to the other.

The Adiabatic model has several setbacks. The most important is the lack of a guaranteed fault tolerant method. In other words it is not clear how to control the amount of noise or decoherence embedded into such computers.

Sometimes the adiabatic computer fails although the problem is simple. It was suggested by van Dam \cite{van Dam} that the adiabatic computer behaves as a ``local search'', {\it i.e.} that on a problem space having a large set of metastable states the time complexity could be high. Surely, the algorithm can not stay on a metastable state being adiabatic, however the fact that there are many such states appears as an exponential decrease of the first gap between eigenvalues. We can therefore find simple problems which take an exponential time to be solved on an adiabatic computer \cite{van Dam}. It was also claimed that AQC can suffer from Anderson localization \cite{Altshuler}.

As for the relation between the first gap and the temperature, surely if $k_bT$ is much smaller than the gap then the adiabatic evolution will overcome thermal noise. In general, if $k_bT$ is larger than the gap it might be useful to describe the evolution of the Hamiltonian within the context of the master equation \cite{Lidar-Master}.

\section{Quantum Annealing}
\label{ECBTsec:4}

\begin{figure}
\center
\resizebox{0.3\columnwidth}{!}{
 \includegraphics{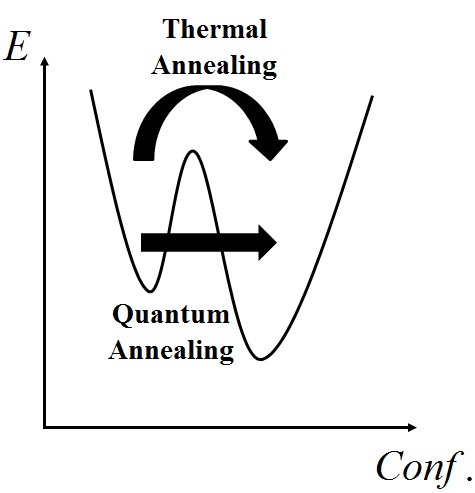} }
\caption{Quantum annealing vs. thermal annealing in a graph of energy as a function of configuration space}
\label{ECBTfig2}
\end{figure}
Quantum annealing \cite{Apolloni,Finnila,Ray} employs quantum fluctuations in frustrated systems or
networks to anneal the system down to its ground state or to its minimum cost state,
and eventually tuning the quantum fluctuations down to zero (see Fig. \ref{ECBTfig2}). By utilizing tunable
quantum fluctuations, this method can be more effective in solving multi-variable
optimization problems, in comparison to classical annealing. The effectiveness
is derived from the fact that unlike classical annealing, where the system
climbs the individual barrier heights by utilizing thermal fluctuations, in quantum annealing fluctuations can help tunneling through these barriers and avoid local minima. Recall that a particle of mass $m$ having energy $E$ will tunnel through a barrier of height $V>E$, with probability amplitude $e^ {-Gx}$, where $G=\sqrt{2m(V-E)}$, and $x$ is the penetration distance \cite{Razavy}. Now since the energy barrier for such spin glass models scales as $N$, the barrier transmission probability will be proportional to $e^{-\sqrt{N}}$ in the quantum annealing case and proportional to $e^{-N}$ in the classical case. Thereby the quantum case is preferred \cite{Mukherjee}.

The method of Quantum annealing was first presented in 1989 \cite{Ray}. A similar technique was suggested for a broader class of continuous problems which was later applied to
configurations of Lennard-Jones clusters and other problems \cite{Johnson,Suzuki}.
Apart from the theoretical research this method has been vastly demonstrated
experimentally, {\it e.g.} in Refs. \cite{Johnson} and \cite{Brooke}.
The quantum annealing method provides an algorithm for combinatorial optimization
problems represented by random models, such as the Ising model. The
process of quantum annealing is realized in principle by the real-time adiabatic
evolution from the quantum initial ground state to the classical final ground state,
the solution to the problem. However there are certain versions where the quantum
annealing protocol is somewhat different: the evolution is not necessarily required to be adiabatic, that is, the system may leave the ground state due to thermal or non-adiabatic evolution (see for example the D-wave protocols below).

Quantum annealing was suggested as an improvement of the simulated annealing
technique which suffers a severe setback in cases where the system is ``non-ergodic''
({\it e.g.} systems described by the spin glass model). In such cases, configurations of $N$
spins corresponding to minimum of the cost function could be separated by $O(N)$
sized barriers \cite{Moore}, so that at any finite temperature thermal fluctuations takes practically infinite time to relax the system to the global minimum.

There are clear similarities between simulated and quantum annealing. In both methods, one
has to strictly control the relevant parameters and change them slowly to tune the
strengths of thermal or quantum fluctuations. In addition, the main idea behind
both classical and quantum annealing is to keep the system close to its instantaneous ground state. Quantum annealing excels in tunneling through narrow (possibly cuspidal) barriers. Classical simulated annealing schedules might still have an advantage where the barrier is wide and shallow.

The basic scheme is as follows. First the computational problem has to be mapped to a corresponding
physical problem, where the cost function is represented by some Hamiltonian $H_0$ of the Ising form:
\begin{eqnarray} \label{ECBTIsing}
H_0=-\sum_{i<j}J_{ij}\sigma_i^z\sigma_j^z-\sum_ih_i\sigma_i^z,
\end{eqnarray}
where $J_{ij}$ denotes the interaction strength between spins $i$ and $j$, and $h_i$ describes the
magnetic field at site $i$. Then a suitably chosen non-commuting quantum tunneling
Hamiltonian $H_1$ is to be added, so that the total Hamiltonian takes the form of:
\begin{eqnarray}
H=H_0-\Gamma(t)\sum_i\Delta_i\sigma_i^x\doteq H_0+H_1(t),
\end{eqnarray}
where $\Delta_i$ denotes the interaction strength with the ``tunneling'' term and $\Gamma(t)$ describes its
time dependence. One can then solve the time dependent Schr\"{o}dinger equation for the
wave-function describing the lattice $\psi(t)$:
\begin{eqnarray}
i\hbar\frac{\partial \psi}{\partial t}=(H_0+H_1)\psi.
\end{eqnarray}

The solution approximately describes a tunneling dynamics of the system between
different eigenstates of $H_0$. Like thermal fluctuations in classical simulated annealing,
the quantum fluctuations owing to $H_1(t)$ help the system to come out of the local
``trapped'' states. Eventually $H_1(t)\rightarrow 0$ for $t\rightarrow 1$ and the system settles in one of the
eigenstates of $H_0$; hopefully the ground state. This serves as a quantum analog of cooling
the system. The introduction of such a quantum tunneling is supposed to make the high
(but very narrow) barriers transparent to the system, and it can make transitions to
different configurations trapped between such barriers, in course of annealing. In other
words, it is expected that application of a quantum tunneling term will make the free
energy landscape ergodic, and the system will consequently be able to visit any configuration with finite probability. Finally the quantum tunneling term is tuned to zero to get back the Ising Hamiltonian. It may be noted that the success of quantum annealing is directly connected to the replica symmetry restoration in quantum spin
glass due to tunneling through barriers.

What are the differences between simulated and quantum annealing? A convergence criterion was proved in \cite{Morita} which is similar to the one in simulated annealing. If we let $\Gamma (t)= t^{-\gamma/ N}$, (where $\gamma$ is some positive constant) we are guaranteed to get a solution. This, however, could take an infinite amount of time. If we stop the relaxation at some final time $t_f$ where the ``temperature'' $\Gamma (t)$ is small, $\Gamma (t)= \delta$ then it is enough to wait $t_f=e^ {N ln(\delta)/2\gamma}$. If we compare this to the relaxation time $t_f$ for the simulated annealing protocol we find that for very small $\delta$, {\it i.e.} $\frac{1}{\delta} >> ln(\delta)$, the quantum annealing scheme will be better than its simulated annealing counterpart. This is true in general, but could be hard to utilize, since both relaxation times are exponential.

In fact, for some specific problems the advantage of quantum annealing over simulated annealing is much more clear. In \cite{Kadowaki} it was tested on a toy model of 8 qubits with a transverse Ising field. The authors showed that quantum annealing leads to the ground state with much larger probability than the classical scheme, when the same annealing schedule is used. In \cite{Martonak} path-integral MC quantum annealing showed better results for the Traveling Salesman Problem for 1002 cities. Here the algorithm was stopped after various number of steps and the results were compared to a simulated annealing algorithm. QA was shown to anneal more efficiently, and to decrease the solution residual error at a much steeper rate than SA. The authors in \cite{Farhi2} constructs an example where the width between local minima is small and therefore the tunneling effect is strong. The simulated annealing counterpart of the example shows an exponential complexity.

Recent results suggest that for first order phase transitions the adiabatic algorithm has exponential time complexity. For second order phase transition the adiabatic algorithm has polynomial complexity. It was also suggested that by adding an annealing term one can solve the first order transition problems \cite{Jorg,Seoane}.

Brooke {\it et al.} \cite{Brooke} applied the above model to a disordered ferromagnet. Their aim was to find the
ground state for the ferromagnet with a certain proportion of randomly inserted antiferromagnetic
bonds. Cooling it down to 30 mK and varying a transverse magnetic field, they were able to compare simulated and quantum annealing, concluding eventually that their experiment directly demonstrates the power of a quantum tunneling term in the Hamiltonian.

The quantum annealing process can be also simulated by a Quantum MC techniques. These basically use the above classical MC techniques to evaluate the multi-dimensional integrals that arise when solving quantum many-body problems. Some of the common methods are the Variational \cite{McMillan}, Diffusion \cite{Grimm}, Auxiliary Field MC \cite{Ceperley}, Path Integral \cite{Barker}, Gaussian \cite{Corney}, and Stochastic Green Function \cite{Rousseau}

The quantum annealers discussed above are all adiabatic. There are deep connections between the behavior of the ``first gap'', the Hamming distance between the ground state and the first excited state, and the transverse tunneling vector. If this Hamming distance near the anti-crossing is $d$ then to cross that distance by tunneling is $\Gamma^d$ (exponentially) hard (this is clear if we look at the transverse field as perturbation). Therefore, if the evolution is too fast we will end (get stuck) at an exited eigenvector with big Hamming distance from the true solution. Hence, hard problems (NP) will either take exponential time or give results with high error \cite{Boixo2}.

There is also a deep connection between the number of free qubits in the $0$ and the first excited eigenstates and the first gap. It is easy to see that the transfer field breaks the degeneracy of a free qubit. Thus, if the first excited state has more free qubits than the $0$ eigenstate, the splitting of the energy states by the transverse field are such that the minimal gap reduces. This does not happen in classical simulated annealing. Therefore, such problems could be harder for quantum simulated annealing than for classical annealing \cite{Boixo2}.

\section{Physical Implementation}
\label{ECBTsec:5}

The D-Wave computer makes use of flux qubits. These are superconducting circuits with several JJs which are also known as SQUIDs (superconducting quantum interference devices). Groups of such qubits are clustered into a cell unit. The first version of the D-Wave contained 128 qubits, the current computer holds 512 such qubits. In what follows we describe the main principles of these circuits \cite{Wendin,Martinis}.

\subsection{Josephson Junctions}
\label{ECBTsec:5.1}
A JJ consists of two superconductors coupled by a weak link. We will present the construction of qubits out of configurations of JJs. All such configurations are described by Hamiltonians of two quantum conjugate variables. The first is the charge or the number of Cooper pairs, the second is the phase shift between the two leads of the JJ. We can use either of the two variables to present a ``charged'' qubit or a ``flux'' qubit. We can describe the qubit by observing the potential function in the Lagrangian of the system. The distance between energy levels in the potential must not be a constant. This way, we can use the lower energy eigenstates as a qubit, and distinguish these lower states from the rest, which will make it easy to manipulate and measure the qubits. In the following we concentrate on flux qubits, these are used in the D-Wave system. \\

The current through a JJ is given by:
\begin{eqnarray}
I_J = I_C \sin(\phi),
\end{eqnarray}
\noindent where $I_C$ is the critical current of the junction and $\phi$ is the phase difference across the junction. In case a voltage is applied to the junction we also have:
\begin{eqnarray}
\frac{d\phi}{dt} = \frac{2C}{\hbar} V.
\end{eqnarray}
\noindent Consider a loop with a JJ. If a (perpendicular) magnetic flux $\Phi$ is applied through the loop then the phase difference across the junction satisfies (mod $(2 \pi)$):
\begin{eqnarray}
\phi = \frac{2e}{\hbar} \Phi = \frac{\Phi}{\Phi_0}
\end{eqnarray}
\noindent where $\Phi_0 = \frac{\hbar}{2e}$ is the magnetic flux quantum.

\subsection{Current based JJ}
\label{ECBTsec:5.2}

\begin{figure}
\center
\resizebox{0.5\columnwidth}{!}{
 \includegraphics{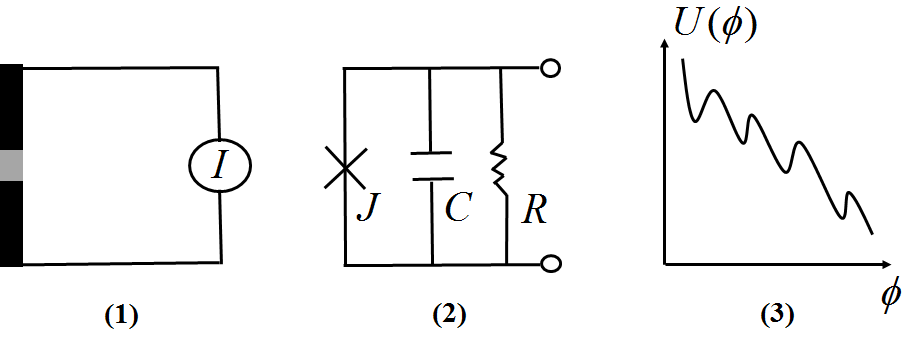} }
\caption{Current based JJ}
\label{ECBTfig3}
\end{figure}

Consider now a JJ biased by an external current $I_e$. If $C$ denotes its internal capacitance, $R$ its dissipative internal resistance, then the phase $\phi$ satisfies the following differential equation:
\begin{eqnarray}
\frac{\hbar}{2e} C \ddot{\phi} +  \frac{\hbar}{2eR}\dot{\phi} + I_c \sin(\phi) = I_e.
\end{eqnarray}

We can assume the dissipation is low and therefore the expression with $R$ can be dropped. We can then write a Lagrangian for the variables $\phi$ and $\dot{\phi}$ such that the above formula is its equation of motion:
\begin{eqnarray}
L(\phi, \dot{\phi}) = \frac{\hbar^2 \dot{\phi}^2}{4E_C} - E_J(1-\cos(\phi)) + \frac{\hbar}{2e} I_e \phi,
\end{eqnarray}
\noindent where $E_C = \frac{(2e)^2}{2C}$ is the charging energy of the junction capacitor, and $E_J =\frac{\hbar}{2e} I_C$ is the Josephson energy. Note that the potential of the above Lagrangian is:
\begin{eqnarray}
E_J(1-\cos(\phi)) - \frac{\hbar}{2e} I_e \phi.
\end{eqnarray}

\noindent This is known as the ``washboard potential'' (see Fig. \ref{ECBTfig3}). \\

One can use the two lowest energy states as a qubit. For a big enough $I_e$, the washboard potential is tilted (Fig. \ref{ECBTfig3}). This can be used to read out the state as discussed below. It is important to note here that since the potential is based on a cosine function the distance between the energy levels is not constant. This makes it easy to manipulate the state, therefore to read and write into the junction.

\subsection{RF-SQUID}
\label{ECBTsec:5.3}

\begin{figure}
\center
\resizebox{0.4\columnwidth}{!}{
 \includegraphics{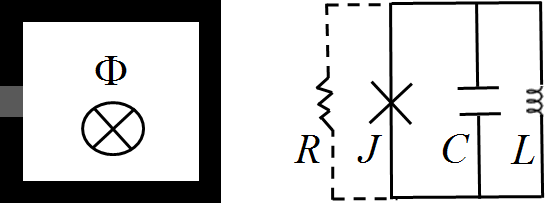} }
\caption{RF-SQUID}
\label{ECBTfig4}
\end{figure}

If we insert a JJ into a loop we actually add an induction $L$ to the circuit (see Fig. $\ref{ECBTfig4}$). Suppose a magnetic flux is induced through the loop. Writing now the potential energy of the variable $\phi$ we get:
\begin{eqnarray}
U(\phi) = E_J (1-\cos \phi) + E_L \frac{(\phi-\phi_e)^2}{2},
\end{eqnarray}

\noindent where $\phi_e = \frac{2e}{\hbar} \Phi_e$ and  $E_L = \frac{\Phi_0^2}{4\pi^2 L}$. If we change variables: $\tilde{\phi} = \phi-\pi$ and $\delta = \phi_e - \pi$ then (up to a constant factor) we have:
\begin{eqnarray}
U(\tilde{\phi}) = E_L( -\epsilon \frac{ \tilde{\phi}^2}{2} -\delta \tilde{\phi} + \frac{1+\epsilon}{4!} \tilde{\phi}^4)
\end{eqnarray}
\noindent where $\epsilon= \frac{E_J}{E_L} -1$. Looking at the above potential we see an inverted parabola in $\tilde{\phi}$ and a 4th power in $\tilde{\phi}$. These two give a double well function where the barrier height is $\epsilon$. The $\delta \tilde{\phi}$ factor which is fixed by an external flux determines the difference between the two minima. Therefore we can fully control the landscape of the potential. We can lower the barrier height to allow the tunneling. The phase will tunnel through the barrier to form a superposition of two states, corresponding to the two opposite directions of the circulating current in the loop. We will get two such eigenvectors having an energy gap proportional to $\sqrt{\delta^2 + \Delta^2}$ where $\Delta$ is a tunneling coefficient in the restriction of the Hamiltonian of the SQUID to its two lowest energy eigenstates. We can use such RF-SQUID as a flux qubit.

\subsection{Persistent Current Qubit, PCQD}
\label{ECBTsec:5.4}

\begin{figure}
\center
\resizebox{0.3\columnwidth}{!}{
 \includegraphics{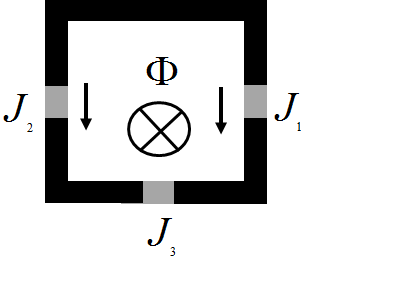} }
\caption{PCQ-SQUID}
\label{ECBTfig5}
\end{figure}
The PCQ consists of 3 junctions (see Fig. $\ref{ECBTfig5}$). These are inserted into a loop. An external magnetic flux $\phi_e$ is driven through the loop. We want the qubit to have two circulating currents of opposite signs. The flux quantization around the loop should satisfy: $\phi_1 -\phi_2 +\phi_3 = -2\pi \phi_e$ (where $\phi_e$ is the external flux). Then the potential energy function is a function of several phase shifts:
\begin{eqnarray}
\end{eqnarray}
\[\frac{U (\phi_1, \phi_2)}{E_J} =  2+ \alpha - \cos \phi_1 - \cos \phi_2 - \alpha \cos(2\pi \phi_e+ \phi_1 - \phi_2),\]

\noindent where $E_j = E_{J_1}=E_{J_2}$ and $\alpha =E_{J_3}/E_{J} $ \cite{Orlando}. One can pick an axis in the $(\phi_1, \phi_2)$ phase space on which the system behaves as if under a double well potential.

\subsection{Reading the qubit state}
\label{ECBTsec:5.5}

Reading the current based single JJ qubit is done by tilting the washboard potential by increasing the external current. The higher energy state can then tunnel through the barrier and drift into a distant value. A big change of the phase could be easily read as a voltage over the junction. The tilt is not enough for the lowest energy state and therefore one can distinguish between the two states.

Reading a flux qubit can be done by coupling the SQUID to an RF oscillator. The oscillator's frequency will change depending on the state of the qubit. This read out technique is similar to the one used in NMR.

\subsection{Coupling the qubits}
\label{ECBTsec:5.6}

Coupling of two systems is done using the external variables of the Lagrangians. For example in an RF-SQUID, $\phi_e$, the external flux of one RF-SQUID, can be the flux induced by a neighboring RF-SQUID loop. We can ``weave'' one of the RF-SQUIDs over the other.
Charged qubits could be coupled by a capacitor. These coupling methods are the ones used in electronic circuits.

\subsection{The D-Wave structure}
\label{ECBTsec:5.7}

\begin{figure}
\center
\resizebox{0.5\columnwidth}{!}{
 \includegraphics{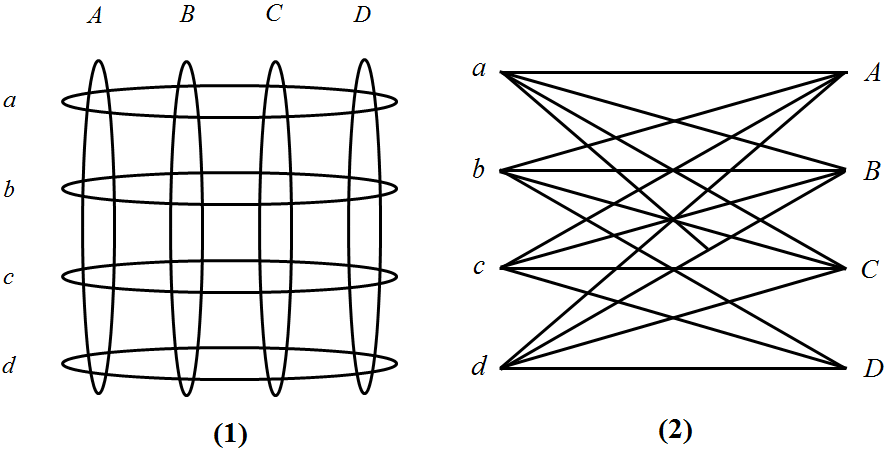} }
\caption{8 qubit cell}
\label{ECBTfig6}
\end{figure}

The D-Wave computers utilize flux qubits of the PCQ type (see above). A set of 8 qubits inter-couple into a cell. In Fig. $\ref{ECBTfig6}$ qubit $a$ is coupled to qubits A,B,C and D. Similarly, qubit A is coupled to qubits a,b,c and d. All 8 qubits and their interconnections can be described by the graph in the figure. In D-Wave Two, 64 such cells constitute a two dimensional grid. Each cell is connected to its neighboring cells. The whole 512 qubits therefore implement a graph known as the Chimera graph $C_n$.

Since $C_n$ is not a complete graph it is not clear how to implement a general graph $G$ into the hardware. One should distinguish between logical qubits and physical ones. Suppose some vertex $v_i$ of $G$ has degree $k$, then we need to couple the vertex (qubit) to $k$ other vertices (qubits). This could be done by mapping $v_i$ into a subtree of several vertices (qubits), we will use the leaves of the tree to connect to other vertices. On the hardware we will get a graph $\mathcal{G}$. The original graph $G$ is called the Minor of the expanded graph $\mathcal{G}$. Such a simple embedding demands that $|G|^2 \sim |\mathcal{G}|$, however there are more efficient embeddings (the task of finding such an embedding is in itself a hard computational problem, see also \cite{Choi}).

The qubits (spins) are coupled together using programmable elements which provide an energy term that is continuously tunable between ferromagnetic and anti-ferromagnetic coupling, this allows spins to favor alignment or anti-alignment, respectively. The behavior of this system can be described with an Ising model Hamiltonian, similar to the one in Eq. \ref{ECBTIsing} above.
During quantum annealing, the transverse terms are gradually turned off and the weight of the Ising Hamiltonian $H_0$ is increased. If this annealing is done slowly enough (according to Eq. \ref{ECBTadiabaticT}) the
system should remain in the ground state at all times due to the adiabatic theorem discussed above, thus ending up in the ground state of $H_0$.

The whole computer is refrigerated to 20 mK by a dilution refrigerator, the pressure is set to $10^{-10}$ atmospheric pressure, and the computer is shielded magnetically to $5 \cdot 10^{-5}$ of earth magnetic field. The D-Wave computer has the size of a small chamber, however its core 512 qubits board is much smaller.

\section{Is the D-Wave a Quantum Computer? The Future of Quantum Annealers}
\label{ECBTsec:6}
The ``Orion'' prototype of the D-Wave contained only 16 qubits. Some of the problems it solved were pattern matching, seating arrangement and a Sodoku puzzle \cite{Sodoku}. The ``D-Wave One'' version incorporated 128 qubits. Using it, a team of Harvard University researchers presented results of the largest protein folding problem solved to date \cite{Ortiz}. The current ``D-Wave Two'' consists of 512 qubits  (not all active) and thus enables to solve much more complex problems like network optimization and radiotherapy optimization which were demonstrated by the company \cite{Dwave}.

In \cite{McGeoch} it was claimed that on Quadratic Unconstrained Binary Optimization problems the D-Wave hardware returns results faster than the best known IBM CPLEX Optimizer, by a factor of about $3600$ for a problem of size $N=439$. This was contested by CPLEX developers in \cite{JeanFrancois}. Several similar claims were recently made by the D-Wave group, but for each such claim there is a contesting claim questioning the results.

The D-Wave computer is presented as a quantum annealer, or as an adiabatic computer having a programmable transverse field for tunneling. In general, it is hard to guarantee that the time evolution will meet the requirements of the adiabatic theorem, and indeed it turns out that the D-Wave computer is manifesting a regime which is in-between adiabatic and thermic. The excitations to higher eigenvectors in the course of evolution is expected to be followed by a later relaxation into a ground state. It was claimed by the D-Wave group that such a regime could improve the probability to get the correct result \cite{Dickson}.

As a new apparatus it is only natural to ask how can we be sure this machine is indeed a quantum computer. In what follows we present several criteria which we believe are important for the identification of the D-Wave (or any other computer) as a quantum computer. Some of the criteria can be attributed to the D-Wave computer, a few cannot, and the rest are still in controversial.

\subsection{Universality}
\label{ECBTsec:6.1}

For the adiabatic model there is no natural set of universal computers generating the whole theoretic spectrum of the model. What kind of problems can the D-Wave computer solve? Currently, only particular optimization problems are considered. However, as was shown in \cite{Farhi} adiabatic computation can solve the 3-SAT problem (in exponential time) and therefore in principle any NP hard problem could be solved. The (polynomial) equivalence of adiabatic computation and quantum circuit computation \cite{Aharonovd} suggests that, theoretically, the adiabatic method can be generalized towards solving any problem that can be solved by the circuit model. Indeed, several protocols for solving specific problems other than optimization were suggested: Graph Isomorphism, Quantum Counting, Grover's search problem, the Deutsch-Jozsa problem and Simon's problem \cite{Farhi,van Dam,Hen3,Hen4}. However, in practice, there is no direct and simple way to translate a solution in terms of the circuit model to solution in terms of adiabatic computation (then again, see the method suggested by Lloyd in \cite{SLloyd}). This is mainly due to the fact that the exponentiation of a sum of Hamiltonians that do not commute is not the product of the exponentiation of each of the Hamiltonians. Therefore, in the circuit model we can present a simple set of universal gates, whereas for the adiabatic model it is much harder.

\subsection{Coherence time of the SQUIDs}
\label{ECBTsec:6.2}

The coherence time of the qubit should be larger than the time needed for the algorithm to compute. This is far from being achieved in the D-Wave computer. The coherence time of the SQUID is about 10 ns while the annealing time needed is 5-15 $\mu$s \cite{Boixo2}. Indeed, how can one achieve quantum computation when the annealing time (depending on the first energy gap) of the computer is about 3 orders of magnitude longer than the predicted single-qubit decoherence? This seems to force a thermodynamic regime on the computer. The main reason for using such flux qubits is that they are relatively simple to manufacture, using common methods of lithography. A set of such SQUIDs, their coupling apparatus and measurement gates are concatenated as on a printed circuit board. Hence, we are still in need for the qubit -``transistor'', that is, a simple apparatus presenting a behavior of a two-state system, that can maintain coherence for a long time (in comparison to the operation time), and can be read off, coupled, and easily manipulated.

The D-Wave computer operates under a semi-adiabatic-semi-thermodynamic protocol. Near the anti-crossing the evolution might be too fast for an adiabatic computer. Therefore, the eigenvectors are excited to a higher energy states, later to be relaxed to the $0$ eigenstate again. This could be done without worrying about the phases, because the ground state is invariant \cite{Farhi}.

As for the question of thermodynamic relaxation, it is suggested by the D-Wave group that such relaxations could only help in the evolution into the ground state \cite{Dickson}.

\subsection{Scalability}
\label{ECBTsec:6.3}

How many qubits can a D-Wave have? D-Wave computers have made a major leap when incorporating the largest number of qubits ever seen on a single device. The question now is of scalability. It is possible that the complexity to construct such a computer with all its inter-couplings, grows itself exponentially. This will mean that the possible gain in algorithmic complexity is paid out in building a coherent circuit (see also \cite{Kalay}). This question is deeply connected to the lack of fault tolerant gate theory for adiabatic computation. A scalable architecture of adiabatic computing was suggested in \cite{Kaminsky}, by translating NP hard problems to the Max Independent Set problem. For that problem a highly robust Hamiltonian was suggested. A more fundamental research in this issue should be done in the context of the master equation (see also \cite{Albash}).

\subsection{Speed-up}
\label{ECBTsec:6.4}

Are the D-Wave computers faster than other computers running different optimization algorithms? For which problems? Right now, it seems that the answer to the first question is ``sometimes'' and it is not clear enough what is the answer to the second problem \cite{Lidar}. In our opinion this is the most important indicator because of its practical significance, but currently it is a problematic issue \cite{Lidar,Katzgraber}. In 2013 it was indeed admitted by the D-Wave group \cite{d-wave1} that different software packages running on a single core desktop computer can solve those same problems as fast or faster than D-Wave's computers (at least 12,000 times faster than the D-Wave for Quadratic Assignment problems, and between 1 and 50 times faster for Quadratic Unconstrained Binary).

In \cite{DefDet} the question of defining and detecting quantum speedup was discussed. Consider the data in \cite{Lidar,newLidar}, where 1000 different spin-glass instances (randomly picked) where investigated. Each instance was run 1000 times and the success probability $s$ for finding the correct solution was computed. The parameter $s$ could also describe the ``hardness'' of the problem.

One annealing run takes $t_a$ time and has a success probability $s$. Therefore the total success probability of finding the solution at least once in $R$ runs is $ p = 1- (1-s)^R$. Set now $p=0.99$ and assume $R=R(s)$. Let $T_{DW}(N,s)$ be the time complexity of the D-Wave computer wired to a problem of size $\sqrt{N}$ (see the above discussion on Minor graphs), and hardness $s$. Clearly $T_{DW}(N,s)$ would be proportional to $R t_a$. One way to define a speedup would be to look at the quotient of quantiles:
\begin{equation}
\frac{T_{DW}(N,s)|_{s\leq s_0}}{T_{C}(N,s)|_{s\leq s_0}}.
\end{equation}
\noindent This means that we average both complexities on a large set of instances (indexed by their hardness $s$) and only then compute the quotient. This suggests a new way to look at computational complexity theory. Another way to define the speedup would be to look at the quantile of the quotient:
\begin{equation}
 \frac{T_{DW}(N,s)}{T_{C}(N,s)}\bigg |_{s\leq s_0}
\end{equation}

\noindent which compares the complexity of both computers on the same instance and only then as a function of hardness $s$. Both methods presented an inconclusive results of speedup, although the second showed a small advantage for using the D-Wave computer when $N$ is high.

\subsection{``Quantumness''}
\label{ECBTsec:6.5}

Since there is no clear evidence for a speedup there is the possibility of comparing the behavior of the D-Wave computer to other models of computation with respect to a large family of computational problems. Consider again the data in  \cite{Lidar,newLidar}. A histogram describing the number of instances for each success probability $s$ was presented. The D-Wave histogram was found to be strongly correlated with quantum annealers rather than classical annealers. Both the D-Wave and the quantum annealer had a bimodal histogram, a large set of problems which are very easy to solve (high success probability) and a large set that are hard to solve (low success probability). The classical simulated annealer had a normal distribution type of histogram with respect to success probability (hardness to solve). This was considered as a proof for the quantumness of the D-Wave machine.

Note that by the above success probability distributions, a problem that is hard for one computer can be easy for the other, while the distribution for the whole ``hardness'' may look the same. This, in itself, questions the interpretation given to the above results.

These conclusions were also criticized by J. Smolin and others \cite{Shin,Smolin1}. It was claimed that the difference between the histograms could be explained out on several grounds. Simulated annealing algorithms start from different initial points each time, while the adiabatic algorithms start from the same point and evolve almost the same each time. Hence, different adiabatic trials naturally show more resemblance. This implies that time scales for the simulated annealing algorithm and for the adiabatic algorithm could not be compared as such. It would be of interest to increase the number of trials given to the SA. This way, one could probably find a good correlation between the simulated annealing and the D-Wave.

In fact in \cite{Shin,Smolin1} a classical simulated annealing model was presented on a set of 2-dimensional vectors, a compass $O(2)$ model. Indeed the model showed a bimodal behavior with respect to success probabilities \cite{Smolin1}. Being a classical model this questions the above results interpretation of the D-Wave computer.

In \cite{Boixo2} the correlation between the success probabilities of solving the same problem instance on any two computers in the set (SQA, DW, SD, SA) was computed. Note that this time each single instance is tested on two computers. High correlation between the DW and SQA was shown. However, in \cite{Shin} similar correlations (even slightly better) were presented between the classical $O(2)$ model and the D-Wave, suggesting a classical behavior of the D-Wave.

In \cite{Boixo1} the D-Wave One was tested on an artificial problem of a set of 8 spins: 4 core spin and 4 ancillae. In that particular case there was a large space of $0$ eigenvectors. The probability distribution of results on the 0 eigenspace was suggested as a proof for the quantumness of the computer. A classical annealer is expected to relax into a solution which has a large number of metastable states in its neighborhood. It is assumed that a classical annealer will quickly find its place into one of these metastable states and thereafter into the 0 eigenvector. An adiabatic quantum computer will behave differently.  The D-Wave One showed the expected adiabatic behavior. In response it was shown in \cite{Shin} that the $O(2)$ classical model of the same problem shows a distribution of $0$ eigenvectors similar to the adiabatic computer, although it is a classical computer.

As another proof for quantumness, the response of the computer to a change in the properties of the flux qubits, was suggested in \cite{Johnson}. For each qubit the thermal fluctuations are proportional to $e^{-\epsilon(U)/k_\beta T}$, where $\epsilon(U)$  is the barrier height (see the above description of PCQ SQUIDs). If we increase $\epsilon(U)$  the thermal fluctuations gradually stop until they freeze out at some freezing time $t_0^c$, such that  $\epsilon(U)(t_0^c)= k_\beta T$. Similarly, the tunneling effects also freeze out when $\epsilon(U)$  is increased above some value  $\epsilon(U) (t_0^q)$.  We expect the freezing time $t_0^c$ of the thermal fluctuation to be linearly dependent on temperature, whereas the tunneling freezing time $t_0^q$ to be independent of temperature.  The authors thereby apparently proved the existence of a tunneling quantum effect.

\subsection{Does the computer exhibit entanglement?}
\label{ECBTsec:6.6}
According to the last work of the D-Wave group the answer is positive \cite{Lanting} with regard to several measures, namely, negativity, concurrence and susceptibility-based witness.

We can explicitly compute a simple model of annealing with two qubits and a transverse field, both the ground and the first exited states turn out to be entangled vectors \cite{Lanting}. The energy gap between the two entangled states decreases with the decrease in the amplitude of the transverse field (by the protocol). If the amplitude is too low the energy gap is less than $k_{\beta} T$ and from that point onward the state becomes a mixed state. Therefore, for such Hamiltonians, if an energy gap is detected above the temperature of the environment we can assume the ground state is an entangled state.

There is a theorem of Smirnov and Amin \cite{Smirnov} that connects the magnetic susceptibility of the adiabatic Hamiltonian with the existence of entangled states. Define the susceptibility of a qubit $i$ to be:

\begin{equation}
\chi_i^\lambda = \frac{\partial \langle \sigma_i^z\rangle}{\partial \lambda}, \
\end{equation}
where $\lambda$ controls the evolution of the Hamiltonian (such as a time parameter). Suppose $\chi_i^\lambda $ and  $\chi_j^\lambda$ are non zero, $J_{ij} \neq 0$, and suppose the evolution is slow enough to reside on the ground state, then at some point in the evolution process the eigenstate is entangled.

Note however, that the existence of entanglement in the process of computation does not guarantee the quantum properties we need from a quantum computer. Therefore this criterion is weak.

\section{Discussion}
\label{ECBTDisc}
The deep connection between statistical physics and combinatorial optimization was successfully utilized over the
years by classical computers running MC simulations such as the Metropolis algorithm,
and later on, simulated annealing techniques. The successor of these methods, namely, quantum annealing,
provides a framework for optimizing properties of various complex systems. In many important cases, this scheme is faster than classical annealing.

The D-Wave group has definitely made a great progress in the field, both on theoretical and practical aspects. However, the D-Wave computer is now at the apex of a controversy. In the following we wish to raise several questions and ideas concerning future research. \hfill\break
(1) Choosing the hardware or the gates of a quantum computer, there are two main factors to be checked: the coherence time and the operation time. There should be a high relation between the two. The superconducting flux circuits of the D-Wave are far from bering the best in that point. For coherence and operation time scales of other qubits see \cite{Nielsen,Ladd}, for achieving long coherence time (0.1 ms) in superconducting qubits see \cite{Rigetti}. \\

(2) The benefits of the flux qubits of the D-Wave are clear: they are easy to build using known techniques of lithography, the flux qubits are easy to couple, etc. However, with respect to other computational properties they are only moderate \cite{Ladd}. \\
(3) If the D-Wave computer has quantum properties and also thermic properties then the best way to analyze its behaviour is by Markovian Master equations (see also \cite{Albash}). \\
(4) In \cite{Katzgraber} it was suggested that the glassy Chimeras of the D-Wave might not be the right test for quantum annealing. It seems that its energy landscape near zero temperature is too simple and does not have significant barriers to tunnel through. This attenuates the properties we want to use in the quantum computation. \\
(5) It could be that the Chimera graph of the D-Wave makes the embedding of graphs into the computer hard. Different wiring of the computer could make it easy to test other problems \cite{Choi}. \\
(6) It could be useful to test a specific problem having a large computational complexity gap between its classical annealing and quantum adiabatic versions. Consider a cuspidal function with very narrow spikes, circulating a global minima, whereas at the outer side of the circle there are some local minima. Such a test function was suggested by \cite{Farhi2}. It was demonstrated there that the time complexity of a classical simulated annealing is exponential due to the height of the spikes, while an adiabatic computer could easily tunnel (that is, in polynomial time) through the spikes if these are narrow enough. \\
(7) It could be useful to simulate other quantum informational tasks besides optimization, and even to test the D-Wave with hard fundamental tasks such as area law behavior etc. \cite{Wolf}.

Achieving these goals might be very challenging, but success will have far-reaching consequences. Efficient solution of optimization problems will be only the start of running many other quantum algorithms. As noted in \cite{Cohen} many other applications such as image and signal processing, quantum measurement and better study of fundamental questions will be possible. Even harder problems such as running various quantum simulations, or alternatively, studies of artificial intelligence \cite{Artificial} are expected to be dealt within these future computers.

\section{Acknowledgements}
\label{ECBTAck}
This work has been supported in part by the Israel Science Foundation Grant No.
1125/10.


\begin{thebibliography}{}

\bibitem{Shor} P. W.Shor, SIAM J. Comput. \textbf{26},(1997) 1484-1509.

\bibitem{Grover} L. K.Grover, in \textit{Proc. 28th Ann ACM Symp. Theory of Computing} (ACM Press, New York 1996) 212-219.

\bibitem{Feynman} R. Feynman, Int. J. Theor. Phy. \textbf {21}, (1982) 467-488.

\bibitem{Lloyd} S. Lloyd, Science \textbf{273}, (1996) 1073-1077.

\bibitem{Dwave} D-Wave Systems Inc. website: http://www.dwavesys.com.

\bibitem{Das} A. Das, B. K. Chakrabarti, Rev. Mod. Phys. \textbf{80}, (2008) 1061.

\bibitem{Laarhoven} P. J. M.van Laarhoven, E. H. L. Aarts, \textit{Simulated Annealing: Theory and Applications}, (Reidel Publishing Company, Holland 1987) Chs. 1-2.

\bibitem{Boixo1} S. Boixo, T. Albash, F. Spedalieri, N. Chancellor, D. A. Lidar, Nat. Commun. \textbf{4}, (2013) 2067.

\bibitem{Boixo2} S. Boixo, T. F. R$\phi$nnow, S. V. Isakov, Z. Wang, dD. Wecker, D. A. Lidar, J. M. Martinis, M. Troyer, Nature Phys. \textbf{10}, (2013) 218-224.

\bibitem{Aaronson} MIT Prof. S. Aaronson: http://www.scottaaronson.com.

\bibitem{Cohen} E. Cohen, B. Tamir, Int. J. Quant. Inf. \textbf{12}, (2014) 1430002.

\bibitem{MC1} Introduction to Monte-Carlo Technique (the Computational Science Education Project) http://www.chem.unl.edu./zeng/joy/mclab/mcintro.html.

\bibitem{MC2} Introduction to Monte-Carlo Technique (University of Nebraska-Lincoln) http://www.phy.ornl.gov/csep/CSEP/MC/MC.html.

\bibitem{MC3} Introduction to Monte-Carlo Technique (Brighton WebsLtd) http:// www.brightonwebs.co.uk/montecarlo/concept.aspx.

\bibitem{Tee} G. J. Tee \textit{The Monte-Carlo Method} (Pergamon Press, Great Britain 1966) Ch. 1.

\bibitem{Levin} D. A. Levin, Y. Peres, E. L. Wilmer, \textit{Markov Chain and Mixing Time} (American Mathematical Society, Providence RI 2009).


\bibitem{Metropolis} N. Metropolis, A. W. Rosenbluth, M.N.Rosenbluth, A.H.Teller, E.Teller, J. Chem. Phys. \textbf{21}, (1953) 1087-1092.

\bibitem{Kirkpatrick} S. Kirkpatrick, C. D. Gelatt, M. P. Vecchi, Science \textbf{220}, (1983) 671-680.

\bibitem{Cerny} V. $\check{C}ern\acute{y}$, J. Optimiz. Theory Appl. \textbf{45}, (1985) 41-51.

\bibitem{Geman} S. Geman, D. Geman, IEEE Trans. Pattern Anal. \textbf{6}, (1984) 721-741.

\bibitem{Bertsimas} D. Bertsimas, J. Tsitsiklis, Statist. Sci. \textbf{8}, (1993) 10-15.

\bibitem{Cohen2} E. Cohen, R. Heiman, O. Hadar, in \textit{Proc. SPIE8295}, (SPIE, USA 2012) 82950K.

\bibitem{Cohen3} E. Cohen, M. Carmi, R. Heiman, O. Hadar, A. Cohen, in \textit{Proc. BMSB2013}, (IEEE, 2013).

\bibitem{Barahona} F. Barahona, J. Phys. A \textbf{15}, (1982) 3241.

\bibitem{Ingber1} L. Ingber Math. Comput. Model. \textbf{18}, (1993) 29-57.

\bibitem{Ingber2}L. Ingber Math. Comput. Model. \textbf{12}, (1989) 967-973.

\bibitem{Vasana} A. Vasan, K. S. Raju, Appl. Soft. Comput. \textbf{9}, (2009) 274-281.

\bibitem{Holland} J. H. Holland, \textit{Adaptation in natural and artificial systems}, (Univ. Michigan Press, Oxford, England 1975).

\bibitem{Lin} S. Lin, B. W. Kernighan, Oper.Res. \textbf{21}, (1973) 498-516.

\bibitem{Fogel}  L. J. Fogel, A. J. Owens, M. J. Walsh, \textit{Artificial intelligence through simulated evolution}, (John Wiley, New York 1966).

\bibitem{Robbins} H. Robbins, S. Monro, Ann. Math. Stat. \textbf{22},(1951) 400–407.

\bibitem{Benioff} P. A. Benioff, Int. J. Theoret. Phys. \textbf{21}, (1982) 177-201.

\bibitem{Deutsch1} D. Deutsch, Proc. Roy. Soc. Lond. A \textbf{400}, (1985) 97-117.

\bibitem{Albert} D. Z. Albert, Phys. Lett. A \textbf{98}, (1983) 249-252.

\bibitem{Deutsch2} D. Deutsch, Proc. Roy. Soc. Lond. A \textbf{425}, (1989) 73-90.

\bibitem{Kronecker} G. Hardy, E. Wright, \textit{An Introduction to the Theory of Numbers} (Clarendon Press, Oxford 1979).

\bibitem{Jozsa} D. Deutsch, R. Jozsa, Proc. Roy. Soc. Lond. A \textbf{439}, (1992) 553-558.

\bibitem{Boyer} M. Boyer, G. Brassard, P. Hoyer, A. Tapp, Fortsch. Phys \textbf{46}, (1998) 493-506.

\bibitem{Grover1} L. K. Grover, Phys. Rev. Lett. \textbf{85}, (2000) 1334.

\bibitem{Shor1} P. W. Shor, in \textit{Proc. 35th Ann. Symp. Foundations of Computer Science}, (IEEE, Los Alamitos, CA 1994) 124-134.

\bibitem{Simon} D. Simon, in \textit{Proc.35th Ann.Symp. Foundation of Computer Science}, (IEEE, Los Alamitos, CA 1994) 116-123.

\bibitem{Kitaev} A. Y. Kitaev (1995), arXiv:quant-ph/9511026.

\bibitem{Shor97} P. W. Shor, SIAM J. Comput. \textbf{26}, (1997) 1484-1509.

\bibitem{Jozsa1} R. Jozsa, (1997), arXiv:quant-ph/9707033.

\bibitem{Namiki} M. Namiki, S.Pascazio, H.Nakazato, \textit{Decoherence and Quantum Measurements} (World Scientific, Singapore 1997).

\bibitem{DiVincenzo1} D. P. DiVincenzo, Phys. Rev. A  \textbf{51}, (1995) 1015.

\bibitem{Shor2} P. W. Shor, in \textit{Proc. 37th Ann.Symp.Fundamentals of Computer Science} (IEEE, 1996) 56-65.

\bibitem{Farhi} E. Farhi, J. Goldstone, S. Gutmann, M. Sipser (2000), arXiv:quant-ph/0001106v1.

\bibitem{Aharonovd} D. Aharonov, W. van Dam, J. Kempe, Z. Landau, S. Lloyd, O. Regev, SIAM Rev. \textbf{50}, (2008) 755-787.

\bibitem{DiVincenzo2} D. P. Divincenzo, (2000), arXiv:quant-ph/0002077.

\bibitem{Chuang} I. L. Chuang, Y. Yamamoto, Phys. Rev. A \textbf{52}, (1995) 3489.

\bibitem{Knill} G. J. Knill, R. Laflamme, G. J. Milburn, Nature \textbf{409}, (2001) 46-52.

\bibitem{Nuclear} N. Gershenfeld, I. Chuang, Science \textbf{275}, (1997) 350-356.

\bibitem{Vandersypen} L. M. K.Vandersypen, M. Steffen, G. Breyta, C. S. Yannoni, M. H. Sherwood, I. L. Chuang, Nature \textbf{414}, (2001) 883-887.

\bibitem{Cirac} J. I. Cirac, P.Zoller, Phys. Rev. Lett. \textbf{74}, (1995) 4091.

\bibitem{Brennen} G. K. Brennen {\it et al.}, Phys. Rev. Lett. \textbf{82}, (1999) 1060.

\bibitem{Mooij} J. E. Mooij {\it et al.}, Phys. Rev. Lett. \textbf{83}, (1999) 1036-1039.

\bibitem{Imamog} A. Imamog {\it et al.}, Phys. Rev. Lett. \textbf{83}, (1999) 4204.

\bibitem{Platzman} P. M. Platzman, M. I. Dykman, Science \textbf{284},(1999) 1967-1969.

\bibitem{Childs} A. M. Childs, E. Farhi, J. Preskill, Phys. Rev. A \textbf{65}, (2002) 012322.

\bibitem{Paz} J. P. Paz, W. H. Zurek, Phys. Rev. Lett. \textbf{82}, (1999) 5181.

\bibitem{SLloyd} S. Lloyd, (2008), arXiv:0805.2757.

\bibitem{Fock} M. Born, V. A. Fock, Zeitschrift fur Physik A \textbf{51}, (1928) 165-180.

\bibitem{van Dam} W. van Dam, M. Mosca, U. Vazirani, in \textit{IEEE Conf.Proc. 42th IEEE Symp. Foundation of Computer Science}, (IEEE, 2001) 279-287.

\bibitem{Altshuler} B. Altshuler, H. Krovi, J. Roland, (2009), arXiv:0912.0746.

\bibitem{Lidar-Master} T. Albash, S. Boixo, D. A. Lidar, P. Zanardi, New J. Phys. \textbf{14}, (2012) 123016.

\bibitem{Apolloni} B. Apolloni, C. Caravalho, D. De Falco, Stochastic Processes Appl. \textbf{33}, (1989) 233-244.

\bibitem{Finnila} A. B. Finila, M. A. Gomez, C. Sebenik, C. Stenson, J. D. Doll, Chem. Phys. Lett. \textbf{219},  (1994) 343-348.

\bibitem{Ray} P. Ray, B. K. Chakrabarti, A. Chakrabarti, Phys. Rev. B \textbf{39}, (1989) 11828.

\bibitem{Razavy} M. Razavy, \textit{Quantum Theory of Tunneling},  (Word Scientific, Singapore 2003).

\bibitem{Mukherjee} S. Mukherjee, B. K. Chakrabarti, in \textit{EPJ-ST Discussion and Debate Issue: Quantum Annealing: The fastest route to large scale quantum computation?, Eds. A. Das, S. Suzuki} (2014), arXiv:1408.3262.

\bibitem{Johnson} M. W. Johnson {\it et al.}, Nature \textbf{473}, (2011) 194-198.

\bibitem{Suzuki} S. Suzuki, J. -i. Inoue, B. K. Chakrabarti, \textit{Quantum Ising Phases and Transitions in Transverse Ising Models}, (Springer Verlag, Berlin 2013).

\bibitem{Brooke} J. Brooke, D. Bitko, T. F. Rosenbaum, G. Aeppli, Science \textbf{284}, (1999) 779-781.

\bibitem{Moore} B. Drossel, M. A. Moore, Phys. Rev. B \textbf{70}, (2004) 064412.

\bibitem{Morita} S. Morita, H. Nishimori, J. Phys. A \textbf{39}, (2006) 13903.

\bibitem{Kadowaki} T. Kadowaki, H. Nishimori, Phys. Rev. E \textbf{58} ,(1998) 5355.

\bibitem{Martonak} R. Martonak, G. E. Santoro, E. Tosatti, (2004), arXiv:cond-mat/0402330.

\bibitem{Farhi2} E. Farhi, J. Goldstone, S. Gutmann, (2002) arXiv:quant-ph/0201031v1.

\bibitem{Jorg} T. Jorg, F. Krzakala, G. Semerjian, F. Zamponi, Phys. Rev. Lett. \textbf{104}, (2010) 20720.

\bibitem{Seoane} B. Seoane, H. Nishimori, J. Phys. A \textbf{45}, (2012) 435301.

\bibitem{McMillan} W. L. McMillan, Phys. Rev. \textbf{138}, (1965) A442.

\bibitem{Grimm} R. C. Grimm, R. G. Storer, J. Compu. Phys. \textbf{7}, (1971) 134-156.

\bibitem{Ceperley} D. Ceperley, G. V. Chester, M. H. Kalos, Phys. Rev. B \textbf{16}, (1977) 3081.

 \bibitem{Barker} J. A. Barker, J. Chem.Phys. \textbf{70}, (1979) 2914–2911.

\bibitem{Corney} J. F. Corney, P. D. Drummond, Phys. Rev. Lett. \textbf{93}, (2004) 260401.

\bibitem{Rousseau} V. G. Rousseau, Phys. Rev. E \textbf{77}, (2008) 056705.

\bibitem{Wendin} G. Wendin, V. S. Shumeiko, (2005), arXiv:cond-mat/0508729.

\bibitem{Martinis} J. M. Martinis, K. Osborne, (2008), arXiv:cond-mat/0402415.

\bibitem{Orlando} T. P. Orlando, J. E. Mooij, L. Tian, C. H. van der Wal, L. Levitov, S. Lloyd, J. J. Mazo, (1999), arXiv:cond-mat/9908283.

\bibitem{Choi} V. Choi, (2008), arXiv:0804.4884.

\bibitem{Sodoku} J. R. Minkel,  http://www.scientificamerican.com/article/first-commercial-quantum-computer.

\bibitem{Ortiz} A. P. Ortiz, Alejandro, {\it et al.}, Sci. Rep. \textbf{2}, (2012) 571.

\bibitem{McGeoch} C. C. McGeoch, C.Wang, in \textit{Proc. ACM Int. Conf. Computing Frontiers}, (ACM, New-York 2013) 23.

\bibitem{JeanFrancois} J. F. Puget, D-Wave vs. CPLEX comparison, Part 1, Part 2, Part 3, https://www.ibm.com/developerworks/community.

\bibitem{Dickson} N. G. Dickson {\it et al.}, Nature \textbf{4}, (2013) 1903.

\bibitem{Hen3} I. Hen, J. Phys. A \textbf{47}, (2014) 045305.

\bibitem{Hen4} I. Hen, EPL \textbf{105.5}, (2014) 50005.

\bibitem{Kalay} G. Kalai, (2011), arXiv:1106.0485.

\bibitem{Kaminsky} W. M. Kaminsky, S. Lloyd, (2002), arXiv:quant-ph/0211152.

\bibitem{Albash} T. Albash, S. Boixo, D. A. Lidar, P. Zanardi, (2012), arXiv:1206.4197.

 \bibitem{Lidar}  S. Boixo, T. F. R$\phi$nnow, S. V. Isakov, Z. Wang, dD. Wecker, D. A. Lidar, J. M. Martinis, M. Troyer, Nature Phys. \textbf{10}, (2014) 218-224.

\bibitem{Katzgraber} H. G. Katzgraber, F. Hamze, R. S. Andrist, Phys. Rev. X \textbf{4}, (2014) 021008.

\bibitem{d-wave1} http://www.archduke.org/stuff/d-wave-comment-on-comparison-with-classical-computers

\bibitem{DefDet} T. F. R$\phi$nnow {\it et al.}, (2014) arXiv:1401.2910.

\bibitem{newLidar} W. Vinci {\it et al.}, (2014), arXiv:1403.4228.

\bibitem{Shin} S. W. Shin, G.Smith, J.A.Smolin, U.Vazirani, (2014), arXiv:1401.7087.

\bibitem{Smolin1} J. A. Smolin, G. Smith, (2013), arXiv:1305.4904.

\bibitem{Lanting} T. Lanting, {\it et al.},  Phys. Rev. X \textbf{4}, (2014) 021041.

\bibitem{Smirnov} A. Y. Smirnov, H. M. Amin, (2013), arXiv:1306.6024.

\bibitem{Nielsen} M. A. Nielsen, I. L. Chuang, \textit{Quantum Computation and Quantum Information}, (Cambridge University Pres, Cambridge 2000).

\bibitem{Ladd} T. D. Ladd, F. Jelezko, R. Laflamme, Y. Nakamura, C. Monroe, J. L. O'Brien, Nature \textbf{464}, (2010) 45-53.

\bibitem{Rigetti} C.Rigetti {\it et al.}, Phys. Rev. B \textbf{86}, (2012) 100506.

\bibitem{Wolf}, M. M. Wolf, F. Verstraete, M. B. Hastings, J. I. Cirac, Phys. Rev. Lett. \textbf{100}, (2008) 070502.

\bibitem{Artificial} http://www.nas.nasa.gov/quantum/, https://plus.google.com/+QuantumAILab/posts.

\end{thebibliography}
\end{document}